# A mimetic approach to social influence on Instagram

Hubert Etienne & François Charton

## Abstract

We combine philosophical theories with quantitative analyses of online data to propose a sophisticated approach to social media influencers. Identifying influencers as communication systems emerging from a dialectic interactional process between content creators and in-development audiences, we define them mainly using the composition of their audience and the type of publications they use to communicate. To examine these two parameters, we analyse the audiences of 619 Instagram accounts of French, English, and American influencers and 2,400 of their publications in light of Girard's mimetic theory and McLuhan's media theory. We observe meaningful differences in influencers' profiles, typical audiences, and content type across influencers' classes, supporting the claim that such communication systems are articulated around 'reading contracts' upon which influencers' image is based and from which their influence derives. While the upkeep of their influence relies on them sticking to this contract, we observe that successful influencers shift their content type when growing their audiences and explain the strategies they implement to address this double bind. Different types of contract breaches then lead to distinct outcomes, which we identify by analysing various types of followers' feedback. In mediating social interactions, digital platforms reshape society in various ways; this interdisciplinary study helps understand how the digitalisation of social influencers affects reciprocity and mimetic behaviours.

## Key words

Computational philosophy, Influencers, Instagram, Social comparison,
René Girard, Marshall McLuhan

## Introduction

Social media influencers are key social actors whose actions can often have major political or economic consequences. Their growing impact on collective behaviours justifies efforts to study their social function and the psychological mechanisms explaining their influence. Given their importance, one may be surprised that the literature devoted to them remains dominated by marketing research, which simply aims to match an influencer offer with a promotional demand. To fill this research gap, we propose a new research direction underpinned by a methodology of computational philosophy – a new discipline which aims to deepen the understanding of social phenomena in the digital world using transdisciplinary methods that combine philosophical and quantitative approaches (Etienne 2022).

This article lays down the foundations for a new perspective on influencers, rooted in philosophical theories and grounded in online data analysis. It aims to present results relevant to both disciplines and advance the understanding of influencers and online social interactions using René Girard and Marshall McLuhan's theories; it also aims to enrich these theories by describing how they apply to digital spaces. Influencers are viewed as communication systems emerging from a dialectic interactional process between content creators and groups of followers which progressively gather around them. Thus, influencers are mainly characterised by the composition of their audience and the type of posts they use to communicate with them. To examine these parameters, we present two original typologies, built using data on the audiences of 619 Instagram accounts (owned by French, English and American influencers) and follower feedback to 2,400 posts captured by three signals: likes, comments and reports. The results are interpreted in light of Girard's mimetic theory (1961) and McLuhan's media theory (1964). Combining these two theories offers a solid framework to analyse social interactions in digital environments through human and technological mediations.

Girard's mimetic theory holds that humans do not produce their desires but copy them from others ('mediators' or 'models') via an unconscious mimetic mechanism. Girard describes the process by which humans' mimicry leads to rivalry and violence as admiration for a model turns into violent hostility against a rival, and he describes how rivalries between many rivals converge on a single target: a common enemy against whom the group can reunify (Girard 1972). Corroborated by eminent works in learning theories, development psychology (Meltzoff 1995, Garrels 2011) and cognitive neurosciences (Gallese et al. 1996, Rizzolatti et al. 1996), the mimetic theory fertilised cross-disciplinary research and gave birth to a new branch of neuropsychology (Girard et al. 1978, Girard 2008, Oughourlian 2007 & 2013). The Girardian paradigm offers a promising framework for analysing digital interactions. It sheds new light on individual and collective phenomena, from social comparison to social influence, pressure to conform, and hate speech and bullying, and it allows researchers to specify the cross-effects of these phenomena and articulate them in a coherent general model.

McLuhan's media theory offers a complementary framework to Girard's. McLuhan approaches technologies as media which both extend our senses and have profound sensorial, cognitive, and societal impacts on human interactions. The introduction of the train, the printing press and the TV deeply impacted the way cities are organised, power is distributed and people perceive and interact with each other, ultimately mediating social interactions. Social interactions are different whether they are mediatised by a phone, an electronic mailbox, an online chat, or a virtual reality space, and the way each technology impacts people's perception and interactions is what McLuhan calls the 'message' of a media.

Applying Girard's mimetic theory to online interactions in light of McLuhan's theory requires understanding how the digital and physical worlds affect the mimetic theory's expressions differently. In turn, the analysis of social interactions in digital spaces may also contribute to enriching mimetic theory, for instance by addressing one of its main questions Girard left open, that of the factors at play in the selection of one's mediators, for which he provides little information despite the central role of the mediator in his theory (Rey 2016). Approaching the mimetic theory from a digital perspective led Etienne to consider 'collective attention' as the core of mediator selection and to propose that the relationship between individuals and their model should be conceived as a communication system (Etienne, in review).

The first section introduces a theoretical framework to approach influencers, arguing that it is more suitable for studying influencers than traditional marketing approaches, and accompanies the framework with two original labelling methodologies to support such investigation: a classification of influencers and a typology of posts. The second section confirms the relevance of such a methodology, revealing meaningful differences between influencers' classes in terms of profiles, typical audiences, and content. It also exposes the double bind content creators face when growing their influence – keeping their reading contract while changing it – and discusses various strategies influencers implement to address this bind. The third section examines how followers respond to unusual content from various feedback signals. It distinguishes between different types of reading contract breaches and discusses their implications in light of anthropological theories.

# I. Influencers as interactive media

Considering that traditional approaches are unsuitable for understanding the nature and dynamics of social influence on digital platforms, we propose an original perspective on social influencers derived from Girard's mimetic anthropology and McLuhan's media theory. We define an influencer as a communication system emerging from a dialectic interactional process between a content creator and their in-development audience; the system is therefore characterised by the composition of the resulting audience and the type of content the creator publishes to engage their followers. To help study influencers from these two angles, we introduce two original classifications: one for Instagram influencers (CIF) and one for their posts (TIP).

## *Classifying influencers using their fame type and origin*

Existing typologies of social media influencers were developed by marketing specialists to leverage their influence for advertising purposes. To help match an influence offer with a promotional demand, these classifications are mainly based on audience size (Ismail 2018, Penfold 2018, Del Rowe 2018, Business Fishers 2019) and the key topics influencers post about (Reech 2020). As examples of this, the 4Cs (context, consistency, connection, content) and 3Rs (reach, resonance, relevance) frameworks (Solaris & Webber 2012, Kostic et al. 2018) illustrate the marketing approach to influencers, viewed as distribution channels for brands and whose efficiency is captured by performance metrics – especially a reach potential and a capacity to realise this potential through engagement and conversion rates. We believe that these approaches cannot capture the relationship between an influencer and their audience and therefore fail to capture the nature and dynamics of digital influence. As an example, a classification based on audience size tends to only display threshold effects, misclassifying accounts that fall close to the divide between audience groups. It attributes all the dynamics leading to a change in influencer class to a single explanatory factor, a change in the audience, whereas the reverse may be true oftentimes: a change in class results in the growth or shrinking of one's audience. Furtherore, these approaches fail to account for market sizes: in Tab A.2, 'mega-influencers' start at 10M followers in France and in the UK but 50M in the US. More generally, these typologies are tailored to a specific goal – monetizing the influence of social actors – and are built along an external and product-centric perspective. They prove very limited in studying the different aspects of digital influence, such as the relation between influencers and their followers. Because the power of influencers originates from their followers, any study on influence should start from the followers' perspective and not the brands'.

There are other typologies that are more relevant to our ambition: those classifying influencers by their professional activity (Grin 2019), their legitimacy in conveying information (McCleary 2019), or their motivation for posting content (Gross & Wangenheim 2018). All three capture relevant characteristics that distinguish influencers, but they are not holistic enough. Professional activity is not always relevant, especially for small-audience influencers, while celebrities often take on various activities during their career, such as reality TV personality, fashion model, entrepreneur, actor, or singer. Furthermore, their social role cannot be reduced to that of information transmitters, and their motivation, in addition to being speculative and difficult to identify on a large-scale, only relates to one side of the relationship.

In contrast, our approach is rooted in second-order cybernetics, which considers human societies as self-organised systems and tries to model their dynamics. Dupuy (2002) combines this view with Girard's anthropology and thus approaches social groups as autopoietic systems fundamentally characterised by a mimetic dynamic. Since human interactions result from mimicry, the dynamics of a social system cannot be predicted in advance because its attractors, the system's potential final states, and the individuals who drive societal change through mimicry only appear as the system evolves. Following

Dupuy, we approach Instagram as an autopoietic system where mimicry plays a fundamental role in shaping social interactions, which are also conditioned by the platform's design. Through users' interactions and content recommendations, some individuals progressively emerge as mimetic models around whom desires converge and collective attention focuses: digital influencers.

A digital influencer is a representation that emerges from the densification of interactions in a communication space (e.g., a personal page on a digital platform) between the space's administrator and a group of recurring visitors. Through the design of the communication space (all actors may engage with a publication, but only the page's holder can post), the multiplication of interactions strengthens the asymmetric relationship between the administrator and the visitors, resulting in a non-reciprocal mimetic influence which allows the former to 'totalise' the latter as a community. Thus, an 'influencer,' who emerges from this interactive process, should be distinguished from the physical person who administers the page, the 'content creator,' although followers tend to associate them both in a syncretic representation. Kim Kardashian as an influencer, for instance, tends to be closely associated with Kim Kardashian as a person, but Banksy is an influencer on Instagram without being known as an individual, and 'virtual influencers' do not even imply a physical person that embodies them. An influencer thus mediates interactions in two ways. First, they mediate interactions between followers who gathers in a common communication space due to their attraction towards the influencer. In doing so, influencers totalise them as a community by publishing content whose unique function is to trigger interactions. Second, they mediate recursive interactions between themselves as creators and influencers. In growing an influencer persona online, they redefine themselves, changing both the way society perceives them and the way they perceive themselves through the social recognition signals they receive from their followers.

In this perspective, an influencer, just like the Freudian leader of a crowd, does not pre-exist in their community but emerges from it as the product of an interactional system. In a way, influencers are for digital communities what political leaders are for political communities, representing the 'endogenous fixed point' by which a spontaneous collective totalises into a community (Dupuy 1989, Vinolo 2018). This does not mean all followers are equally engaged in the community – one may be an absolute Eminem fan, while another may merely appreciate his music. However, they are all gathered around a common attractor, and we may wonder how attractors come to be in the attention market that is Instagram.

To examine this question, we collected a sample of Instagram accounts from France, the US and the UK with an audience superior to 10K followers. These three countries have the most developed influencer culture (Galaxy marketing, Statistica 2018), and Instagram is the main platform for influencer marketing (Mediakix 2019). An exploratory review of this sample allowed us to build a classification of influencers' fame (CIF) based on two criteria. First, the origin of their fame: either 'organic' for influencers who created a presence purely through digital social media or 'external' for those who were born on another medium – such as TV or magazines – before developing a social media presence. Second, their type of fame: either mainly based on the individual's personality (e.g., reality TV celebrities) or on their professional activity (e.g., sportsmen, artists). Because influencers are distinct from creators, the degree of personification – i.e., how an influencer's personal life and tastes drives their social media presence – is an important factor in assessing how influencers identify with their account holders and how followers identify with them. Tab 1 presents the resulting classification, which we used to label a randomly selected sample S0, extracted from the initial sample, of 619 Instagram accounts. S0 is split into four groups: 152 influencers with more than 10M followers constitute a 'world_stars' category, the 467 others are evenly split between France, the UK and the US ($N_{FR}$ = 154, $N_{US}$ = 156, $N_{UK}$ = 157). The sampling was stratified to ensure a relative balance between influencers' countries, audience ranges and classes.

| Class | Origin of fame | Type of fame | Content personification degree | Content creator' activity |
|---|---|---|---|---|
| 1 | External | Person-based | High | Reality TV personalities |
| 2 | External | Hybrid | Moderate | Fashion models<br>TV presenters |
| 3 | External | Activity-based | Moderate | Actors<br>Singers<br>Sportsmen |
| 4 | Organic | Person-based | Very high | Social media pure players (Tiktokers, bloggers) |
| 5 | Organic | Hybrid | Moderate | Social media pure players (YouTubers, bloggers)<br>Coaches (life, sport, nutrition)<br>Passion brodcasters (wine, hunting, cars) |
| 6 | Organic | Content-based | Very low | Artists (food, tattoos, design, make-up)<br>Passion broadcasters (memes, pets) |

**Tab 1.** Classification of influencers per fame

*Classifying influencers' publications as part of a communication system*

The diversity of influencers and of their associated audiences drives the type of content they post. Following McLuhan, who approaches news magazines as 'mosaics' which reflect the life of a society (1964, 236), we view Instagram as a meta-mosaic composed of various magazines with their own columnists, readerships and reading contracts. Both magazines and Instagram accounts compete for people's attention on an organised market, and new entrants test different strategies until they find the right match between the content they propose and the feedback they receive from subscribers. Although buzz events can suddenly boost an influencer's reach, influence is not built in a day; it results from an interactional process of proposals and adjustments with a growing audience feedback that shapes the influencer's image and influence as their community grows. For example, in France, the YouTubers McFly and Carlito tried several formats before finding one that attracted a community of receptive and loyal followers, shifting from metal music videos to long format videos including jokes and games.

As influencers are shaped by the interactional process they emerge from, which modulates both their representation and their audience's composition, they exist in a communication system characterised by a particular community's expectations about the influencer's communication. The resulting mosaic reflects an influencer's desired images, in accordance with the image their communities expect them to conform with, and every publication is a social interaction feeding this representation. Instagram therefore vividly illustrates McLuhan's assertion that 'the medium is the message' because, with rare exceptions, its posts are not information supports but communication signals. A post's meaning is not to be found in the *message* it conveys ('I drink a cocktail in my swimming pool') but in the *meta-message* it expresses, i.e., what the content says about its creator ('I belong to this successful and cool elite'). Consequently, publications are critical moments when creators test the adequation between the image they believe they are associated with as influencers and the representation through which their followers actually perceive them. They are interactional proposals between creators and their audiences, to which the latter reply in various ways (including likes, comments, reshares, reports, follows or unfollows), and this feedback can be understood as signals indicating how current posts drive audience change.

To investigate the communication signals between different types of influencers and their communities, we developed a typology of influencers' posts (TIP) based on an exploratory review of a sample of content pieces posted by influencers from S0. Tab 2 presents the typology, which combines a pictural label ($TIP_1$), an emotional label ($TIP_2$) and a communicational label ($TIP_3$). We used this typology to label a sample S1 of 2,404 content pieces posted by 204 influencers from S0 over a two-month period (10/31/2021–01/04/2022). After filtering sponsored posts, we were left with 2,142 items balanced between regions and classes and compared the proportion of personified content in each influencer classes and in relation to audience size. Additional information regarding the labelling methodologies and composition of S0 and S1 can be found in Sections 1 and 2 of the Supplementary Materials.

| | Pictural label (1) | | Emotional label (2) | | Communicational label (3) |
|---|---|---|---|---|---|
| 1 | Personal – Me | H | Showy | 1 | Closed – Me |
| | | N | Neutral | | |
| 2 | Personal – Me & others | E | Excitement (announcements, celebrations) | 2 | Closed – Neutral |
| | | P | Proud (self-celebrations, achievements) | | |
| 3 | Personal – Others | S | Self-derision | 3R | Unilateral – Recommendations |
| | | J | Humour (jokes) | | |
| 4 | Professional – Me | L | Love | 3W | Unilateral – Wishes |
| | | M | Motivational (inspirational, spiritual) | 3T | Unilateral – Thanks |
| 5 | Professional – Me & others | G | Gratefulness (or goodwill) | | |
| | | O | Altruism (celebration of someone else) | 4 | Open – Direct message |
| 6 | Professional – Others | C | Charity (or social cause) | | |
| | | X | Sadness (or compassion) | 5 | Open – Contest |
| 7 | Nobody | D | Disgusted (or anger) | | |
| | | A | Authenticity | 6 | Open – Feedback |

**Tab 2.** Typology of influencers' posts (TIP)

### II. Influencers are communication systems organised around a reading contract

If social influence is not to be exhausted by an influencer's reach and publication topics but also captures their history, including the origin of their fame and the type of interactions it have developed with their audience, we should observe meaningful differences between CIF classes in terms of audiences' composition, influencers' content and followers' feedback. The analyses presented here confirm this conjecture, prompting us to approach influencers as communication systems articulated around a 'reading contract' which defines their influence as much as it limits it. While influencers' posts are consistent with their fame type, growing an influence requires them to shift towards more personal content; this presents a challenge for their reading contract, which influencers address via three strategies.

*Different classes of influencers emerge from different types of audiences*

The Girardian theory grants decisive importance to the selection of models in the determination of individuals' identity. It is less that the selection *per se* informs about one's present identity but rather that, mimicking their desires and behaviours, they converge towards their models. Such a distinction is blurred on Instagram, where audiences are composed of both new followers (in the identification phase) and old followers (who have already converged to their model). However, the distribution of influencer and follower age and gender sheds light on the selection process, allowing us to determine whether followers select influencers that resemble them or not. We analyse S0 to answer this question; to neutralise the effect of regional disparities, we only consider national audiences for the three national groups (e.g., French followers of French influencers), but all followers of world_stars.

Figure 1 presents the distribution of influencers' ages and their audiences' age differences per class. In all groups, we observe that two-thirds of followers are less than ten years younger or older than the influencers they follow, i.e., they belong to the same generation. Furthermore, 24% of followers have an age difference of fewer than 3 years with their influencers. These conclusions are confirmed and reinforced by Tab A.4, which presents age differences for each CIF class. Age similarity is strongest for classes 2, 4 and 6 (i.e., age covariance matrices are concentrated on their main diagonal). These observations are less salient for world_stars, probably because audiences are not restricted to national followers. We also note that two-thirds of the followers are younger than the influencers, with the exception of class 4, the youngest influencers. Whereas people follow influencers who are close to them in age, influencers' ages vary between classes, as do audience ages. Class 4 influencers and their followers are the youngest population, with a median age of 25.5. Classes 1, 3 and 5 are slightly older (resp. 31, 32, 31 years old), and classes 6 and 2 have the oldest influencers and followers (resp. medians 34 and 47 years old). Furthermore, influencer age dispersion is lower for classes 1, 4, 5 and 6 than 2 and 3, as is the age difference with their audiences: 40% of the followers of class 1 influencers have less than 5 years difference (54% for class 4, 38% for class 5 and 42% for class 6). These proportions drop to 25% for class 2 and 33% for class 3.

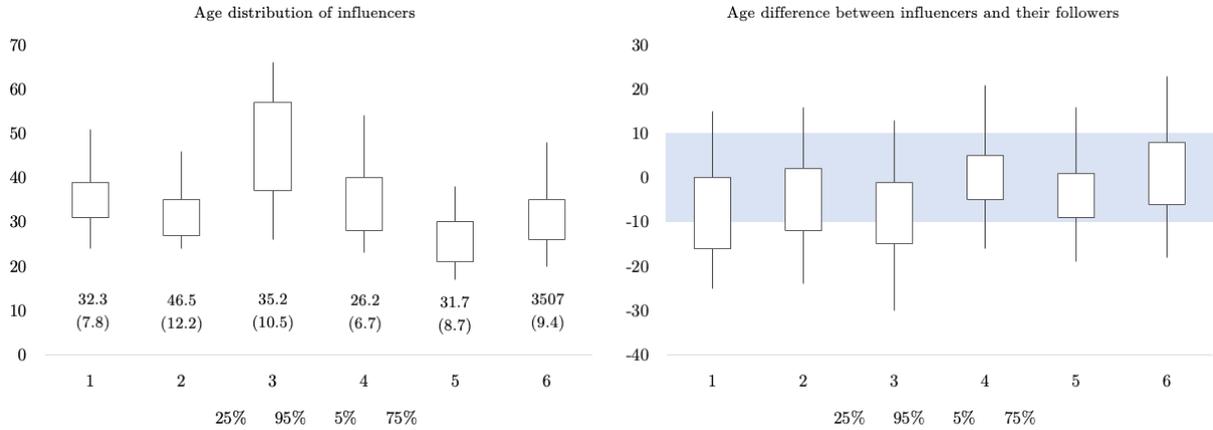

**Figure 1.** Distributions of influencers' ages and their audience's age difference per class

Figure 2 presents the gender split of S0 audiences according to influencer class and gender (on average, audience genders are evenly split along age groups; see Figure A.1). We also compute a gender similarity indicator, which measures the proportion of followers that is of the same gender as the influencer. The closer to 50% this indicator is, the more 'genderless' the influence. Because of their international audiences, results for world_stars are harder to interpret than those of the regional groups. For the regional groups, we observe a significantly higher gender similarity for classes 1, 3, 5 and 6. In these categories, male followers tend to follow male influencers, and female followers female influencers. We note that in classes 1, 2 and 4, associated with influencers whose fame is significantly based on their person, a majority of influencers are female (this is especially true in classes 1 and 4). Followers of these classes are mostly women who follow influencers of both genders. This trend is particularly strong for class 1, where 79% of influencers are female at the national level and 78% of world_stars are female, and class 4 (resp. 85% and 86%); again, these are the two classes of influencers whose fame is the most person-based.

In contrast, in classes 3, 5 and 6, corresponding to influencers primarily famous for their activities, the followers are gendered (i.e., men follow men, women follow women; see Tab A.4). We believe gender similarity results from the gendering of the corresponding activities. Sports audiences, for instance, are known to be gendered, with male sports overrepresented in this study (in class 3, 94% of sports influencers at the national level are men, and 97% of world_stars are men) along with male followers (70% national and 82% world_stars). Passion broadcasters (from classes 5 and 6) display similar gendering, with 83% (male) and 76% (female) similarity among broadcasters. The same is true for social media content creators (64% similarity for men and 77% for women).

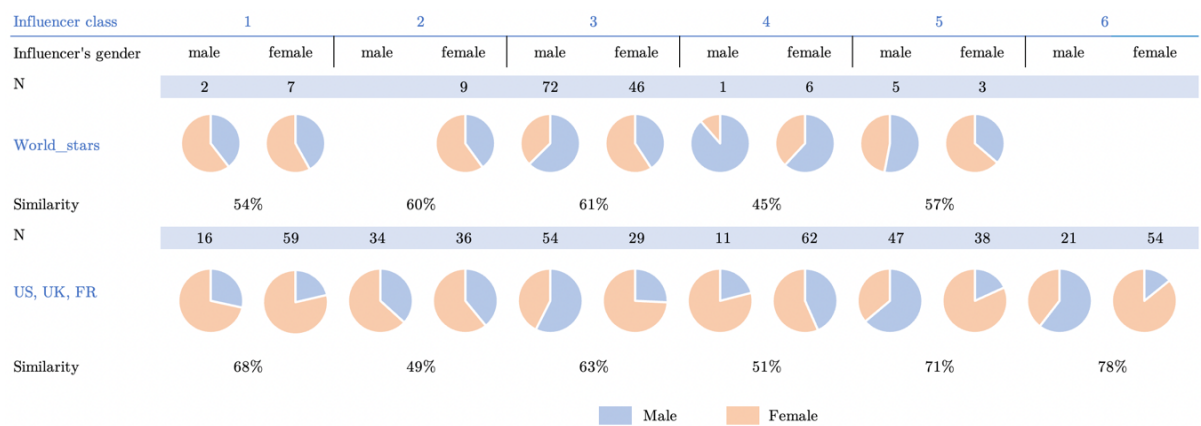

**Figure 2.** S0's audiences' gender split per influencer class and gender

These results confirm the relevance of our classification, which displays meaningful differences between influencers' profiles and audiences across CIF classes while informing us about the selection criteria of social models. Overall, people follow influencers similar in age to them, usually slightly older than them, and the age of followers of different classes correspond to the age of the influencers. Nevertheless, age-sensitivity seems stronger for classes 1, 4, 5 and 6, whereas classes 2 and 3 are more cross-generational. Furthermore, influencers with person-based fame and a high degree of content personification are predominantly women and followed by audiences that are overwhelmingly female, whereas those whose fame is more activity-based and who post hybrid content have gendered followers: men and women tend to follow same-gender influencers, consistently with the gendering of their activities.

As supports for identification and mimicry, Girardian models are highly personified and tend to be of the same gender as their subjects. Classes 1, 4 and 6 could represent different types of female models associated with distinct age ranges: class 4 for women in their early twenties, class 1 for women in their late twenties and class 6 for women in their thirties. Interestingly, while gender similarity increases from class 1 to 4 and to 6, the personification degree decreases dramatically from class 4 to 6, suggesting different types of influence, with class 1 being closer to traditional Girardian models. In the same vein, sportsmen (class 3) and social media content producers (mainly YouTubers in class 5) could represent male-typical models. However, as their fame is more activity-based and these activities appear to follow social gendering, they make less convincing candidates for Girardian models. If typical models appear more clearly for women than for men, it may be because Instagram is used more by the former than the latter in the three selected countries, with 56% of the user population of these countries being women versus 51% for all countries. Previous research has also found that women are more inclined to compare themselves with their peers on social media (Gibbons & Buunk 1999, Guimond 2007, Nesi & Prinstein 2015) and, while Burke et al. (2020) pointed out that these studies were only conducted in Western countries, the countries considered here belong to this category.

*Mimetic influence emerges with a shift in content*

Audience sizes do not sufficiently explain one's influence. However, because audiences develop through interactions between influencers and followers, large audiences constitute a valuable signal attesting to the success of one's strategy in developing influence. In the absence of the historical data needed to study the evolution of influencers' interactions and audiences over a long period, audience sizes within each class can be used as proxies for time when examining the development of an influencer; in this view, an influencer with 100K followers is seen as the 'successful future' of a same-class influencer with 20K followers.

The lower part of Tab 3 presents average audiences per influencer class. Overall, external influencers (classes 1, 2 and 3), whose fame was acquired before they joined social media, have larger audiences than organic influencers (classes 4, 5 and 6). Among the external influencers, person-based fame and a high degree of content personification strongly correlate with large audiences (i.e., class 1 > class 2 > class 3). However, at the world_stars level, there are exceptions: reality TV celebrities (class 1) have the biggest audiences, followed by singers (class 3), actors (class 3), fashion models (class 2) and, lastly, sportsmen (class 3). This hierarchy changes at the regional level: in the US, fashion models and singers have audiences of the same size as reality TV celebrities, in France, sportsmen have the largest audiences, and in the UK, sportsmen and TV presenters are the most followed (see Tab A.3). Yet, at the country level, class 1 (external) influencers dominate audience rankings, and their domination is significantly reinforced once they attain international fame. In contrast, among organic influencers, those with hybrid fame and a moderate degree of content personification have the largest audiences.

One possible explanation for the differences between external and organic audiences is that the audience growth of external influencers is driven by personified content, whereas that of organic influencers is driven by hybrid content. Yet, class 3 influencers owe their fame to their activities, and class 5 influencers may become actors or TV presenters after becoming famous for posting humoristic videos. To further investigate the role of personified content, we compare its proportion between influencers' classes and audience size in S1. Figure 6 presents key indicators reflecting the type of content posted by influencers in S1 and success metrics.

|  |  |  | \multicolumn{6}{c|}{Influencer class} |
|---|---|---|---|---|---|---|---|---|
|  |  |  | 1 | 2 | 3 | 4 | 5 | 6 |
| **Type of content** | **TIP_1** | Self-centred (1) | 28% | 18% | 18% | 59% | 21% | 5% |
|  |  | Personal (1,2,3) | 65% | 46% | 40% | 84% | 42% | 15% |
|  | Audience size | >10M | 68% | 62% | 42% | 85% | 61% | - |
|  |  | 1-10M | 67% | 56% | 34% | - | 71% | 4% |
|  |  | 100K-1M | 62% | 38% | 46% | 84% | 28% | 25% |
|  |  | 10-100K | - | 32% | 26% | 84% | 34% | 14% |
|  | **TIP_2** | Neutral (N) | 28% | 33% | 29% | 34% | 42% | 67% |
|  |  | Top labels (\N) | HL (39%) | JLG (32%) | ELG (35%) | H (44%) | HL (22%) | G (11%) |
|  | **TIP_3** | Close (1-2) | 65% | 68% | 70% | 74% | 54% | 45% |
|  |  | Unilateral (3) | 27% | 25% | 26% | 18% | 26% | 39% |
|  |  | Open (4-5-6) | 8% | 7% | 5% | 7% | 20% | 16% |
| **Success metrics** | Avg. audience (/M followers) |  |  |  |  |  |  |  |
|  | World_stars |  | 130.4 | 40.8 | 48.2 | 22.5 | 22.3 | 10.4 |
|  | Nationals |  | 1.8 | 1.2 | 1.2 | 0.2 | 1.2 | 0.2 |
|  | Avg. likes (/K followers) |  | 29.5 | 25.9 | 40.4 | 60.3 | 31.7 | 16.8 |
|  | Avg. coms (/K followers) |  | 0.18 | 0.54 | 0.77 | 0.80 | 0.87 | 0.47 |
|  | % posts sponsored |  | 15% | 9% | 9% | 15% | 7% | 8% |

**Tab 3.** Summary of S1's type of content per influencer class with their success metrics

Looking at the type of content in S1, we first observe that the proportions of personal content (TIP$_1$⊂(1,2,3)) and of purely self-centred content (TIP$_1$ = 1) match influencers' type of fame for both external and organic influencers. The fame of influencers in classes 1 and 4 is greatly associated with their person, and most of their posts relate to their private life. In class 1, which has influencers with the strongest 'personal brands,' two-thirds of the content is purely self-centred. Instead, for classes 2, 3, and 5, fame is more activity-based, and most of the content is profession-oriented – 54% for class 2, 60% for class 3 and 58% for class 5. Finally, for class 6, influencer fame is entirely activity-based, and 85% of the content is professional, with 72% not even containing an individual's face. Overall, the content posted greatly matches the reasons for the influencers' fame.

Looking at emotion labels ($TIP_2$), we notice that each class can be associated with a narrow set of privileged emotions: $>3/4^{th}$ of the content in class 4 was rated H or N and in class 6 N or G, $>2/3^{rd}$ of the content in classes 1 and 5 was rated H, N or L, and in class 2 N, J, L or G, and in class 3 N, E, L or G.

Finally, patterns also emerge when looking at the communication labels ($TIP_3$): classes 1, 2, 3 and 4 use closed communication in 65-70% of cases, unilateral communication in 25-27% of cases, and almost never resort to open communication. Instead, open communication is a significant part of the posts of classes 5 and 6 (16-20%).

Overall, there is a strong contrast between classes 1 and 4, who mostly post about their private life in a showy way and do not encourage open communication with their audience, and class 6, who almost never advertise themselves, keep a neutral tone and call for dialogue with their followers. The other classes share some of the traits of the two extreme groups. Taken together, these observations support the hypothesis that there exists an implicit 'reading contract' (Veron 1985) between influencers and their followers upon which the communication system is based and which links the influencer class and type of content followers expect them to post. Then, the next question to consider is how this contract evolves as influencers gain fame.

The usual path to notoriety consists in gaining public attention with one activity (e.g., sports, music, YouTube videos, art pieces) to attract an audience. As one's notoriety and audience grow, people are not only interested in the performance but also increasingly want to know about the performer themselves: paparazzi only care about the private lives of people who are already famous. We thus expect influencers' content to evolve as their fame grows from being mainly activity-based to being increasingly focused on themselves and their private life. We call those who follow this pattern 'shifters;' they are represented by classes 2, 3 and 5, for which there is a strong positive correlation between the size of the audience and the proportion of personal posts in relation to professional posts. Instead, for classes 1, 4 and 6, there is no correlation between audience size and proportion of personal content; we call these influencers 'keepers.' Influencers in classes 1 and 4 started by posting personal content and have kept on doing so as their audience has grown. At their antipode, class 6 began by publishing almost exclusively professional content, and this has not varied as their audience has increased.

We interpret these findings as follows: the organic classes (4, 5 and 6), which were all born on social networks, point to three strategies for growing one's influence on Instagram. Class 5 abides by the traditional pattern, progressively shifting from professional to personal content, thus converging, together with classes 2 and 3, their counterparts among external influencers (whose fame is also activity-based). To grow their audience, classes 2, 3 and 5 seek to emulate class 1, which claims the largest audience and posts the most personified content. However, these shifter classes are strongly exposed to a risk of stalling: shifting too quickly from professional to personal content may result in the loss of a significant part of their initial followers because the discrepancy between why the followers engaged with the influencer (their activity) and the content they now publish (mostly personal) becomes too large. Thus, these classes face the significant challenge of shifting from activity-based to person-based fame while ensuring audiences do not stall.

Class 6 illustrates a diametrically opposed strategy. Like class 5, their notoriety is built upon their activity, but they do not try to personify their content to grow their audience. Instead, they keep posting purely de-personified professional content even though this loyalty to their reading contract is detrimental to their reach potential: they claim the lowest audience of all classes and seem to struggle to reach 10M followers. Interestingly, while classes 6 and 1 stand at the antipodes of our typology – they are opposed in fame type and origin, content personification and audience size – they also claim the greatest gender similarity ratios of all classes. This, in addition to the age difference between influencers and audiences in classes 1 and 6, supports the postulate that class 6 includes older women less attracted by social contests and peer comparison. Class 1 manifests 'general influence,' which is pertinent to social mimicry and identification, whereas class 6 exercises 'local influence,' which is specific to a topic. The former is based on unconscious attraction and mimetic desire ('because I am fascinated by X, I want to be more like them and tend to copy everything they do'), the latter on conscious trust ('I trust X about this recommendation in their domain of expertise'). General influence is more powerful because it is orthogonal to the topic or domain of expertise (e.g., Kim Kardashian can support a shoe seller as much as an airline company), but the greater difference dwells on the psychological mechanisms they rely on: one type of follower trusts the judgement of a source of authority in a given field, while the other mimetically copies the desires of a fascinating model of identification. This, supported by the observation that class 1 has the largest audience and class 6 the smallest, prompts us to conclude that the genuine power to influence large populations is necessarily person-based, which is why classes 2, 3 and 5 shift towards personal content as they grow their audience.

Finally, while class 5 follows the path that class 6 refuses, class 4 intends to skip its initial activity-based path to fame. By combining small audiences with greatly personified content, influencers in this class post as if they were already famous and act as if their private life was interesting enough to warrant the fame that usually produces this interest. This strategy reminds of Girard's discussion of 'la coquette,' which Etienne elaborates on to help explain influencers' success on Instagram. Girard defines 'la coquette' as the strategy some characters in novels use: they pretend to be indifferent to others and desire no one but themselves (Girard 1961), and by doing so, they hope that others imitate their desire for themselves. Developing Girard's theories under the prism of attention, Etienne argues that humans are naturally driven by a fundamental desire which leads them to seek objects and individuals worthy of their attention (Etienne, in review), i.e. beings which have traits significantly different from others in a way that make them look superior. He views collective attention – what Instagram captures via audience sizes and engagement metrics – to be the main criteria informing people about each others' value. Facebook-like communities are known in graph theory as small-world graphs (Watts & Strogatz, 1998): tightly knit communities where everyone is connected to everyone loosely linked by a few long-distance relations. Instagram, in contrast, has a scale-free structure (Barabasi & Albert, 1999), where the number of followers forms a hierarchy of audiences, from small-world audiences of regular users who follow their friends to power users with dozens of thousands of followers and super-users with millions (Barabasi 2002). Thus, unlike Facebook, which is a horizontal medium that gathers *friends* who usually already know each other and where friendship is necessarily reciprocal by design, Instagram is a vertical medium where some accounts have thousands of *followers* without reciprocity. Mastering Instagram is not only about having a great number of followers but also about following the lowest number of accounts. In creating this gap, one sends a signal: 'many people are interested in me, but I am not interested in them.' *De facto*, if one does not follow an account, the account does not exist in their feed and, therefore, attention space. Furthermore, Instagram allows for less direct reciprocity mechanisms than Facebook. Facebook posts are invitations for relatively close ties to interact with the author and other commentators: the post is the topic. Conversely, Instagram posts are invitations for followers to interact with the post and among themselves, not with the creator: the influencer is the topic, and followers talk about it but not with it. This justifies the great prevalence of closed communication structures among influencers' posts, especially for external and class 4 influencers. Instagram posts are not designed to initiate bilateral interactions but to produce and maintain an influence through the horizontal interactions of a community of followers.

The coquette strategy is useful for understanding the psychology behind actors' behaviours and the source of their influence. Classes 1 and 4 are the influencers who exhibit her attributes the most, posting self-centred and showy content that invites no interaction. As Dupuy elaborates, the coquette only expresses a 'pseudo-narcissism' (Dupuy 1985, 112) because, despite her pretended autonomy, she needs others to desire her to be able to desire herself; thus, her self-desire is mediated by others' desire. Class 1's reading contract is characterised by 65% of personal content, including 28% of exclusively self-centred ($TIP_1 =1$), mostly showy or neutral ($TIP_2 =$ H (22%) or N (28%)) content, and structured as a closed communication message in 65% of cases ($TIP_3 \subset (1,2)$). Class 4 radicalise this contract with 84% of personal content, including 59% of exclusively self-centred, mostly showy or neutral ($TIP_2 =$ H (44%) or N (34%)) content, and structured as a closed communication message in 74% of cases ($TIP_3 \subset (1,2)$). While the coquette strategy is based on a pretence of autonomy, the illusion appears clearly in S1 because class 4's hypermimetism vis-à-vis class 1 betrays their great attention to others. Finally, Girard asserts that the coquette is an efficient strategy to attract others' attention and desires, which is coherent with our observations. Class 1 dominates in terms of audience size but claims average levels of likes and the lowest amount of comments. In contrast, class 4 records very limited audiences but the highest engagement metrics of all classes. The fact that these two classes make significantly more partnerships than all other influencers is another signal supporting their influential power, which is acknowledged by brands. If coquetry is a relevant concept here, and the strategy efficient, it is because interactions between social media influencers and their audiences are not 'parasocial' as understood by Horton and Wohl (1956) and unlike what Righ and Wegener (2019) argue applies to YouTube influencers – interactions which are 'not mutual, but one-sided.' Instead, followers communicate with the influencer through various means, including likes, comments, reshares, following, unfollowing and even direct messages.

Figure 3 illustrates these three strategies, projecting the proportion of S1 influencers' personal content on their audience sizes for classes 4, 5 and 6, whose number of posts labelled in S1 was superior to 5 (27/30 for class 4, 27/32 for class 5 and 24/36 for class 6). We use a 10-base logarithm scale and the green points represent local averages to reflect the general trends for each of the three strategies.

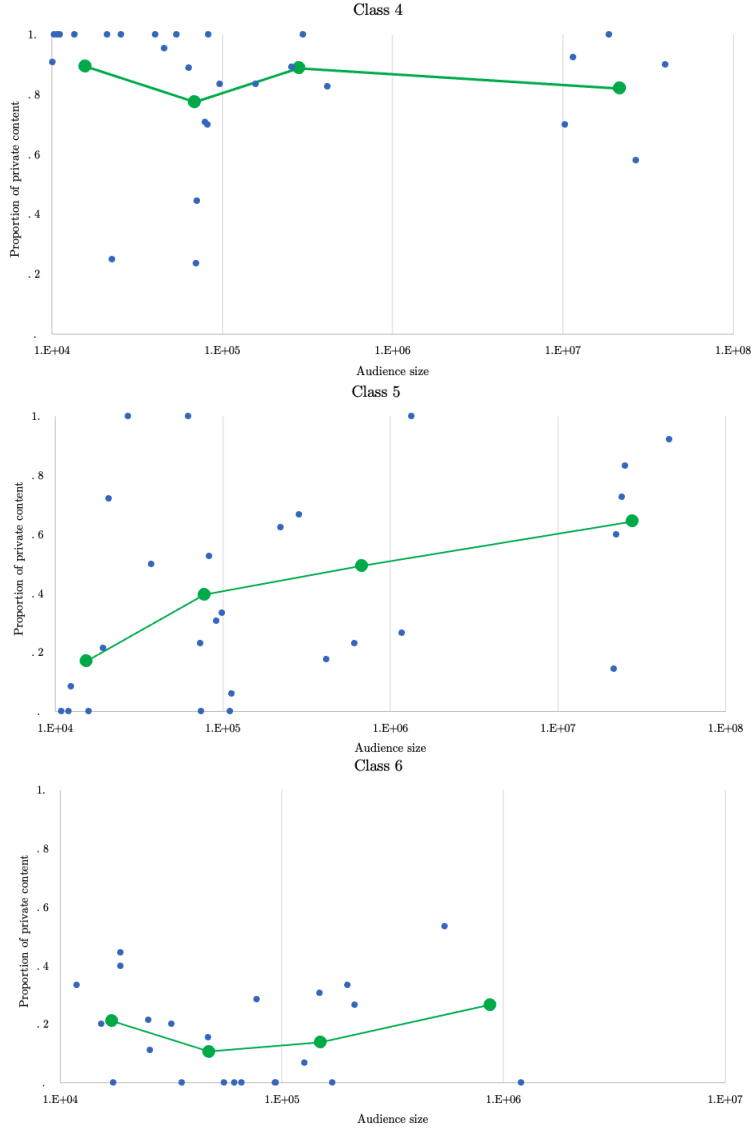

**Figure 3.** Three influencer growth strategies represented by classes 3, 4 and 5.

### III. Publications as critical tests confronting creators and their audiences

The identification of typical audiences and publication styles supports the existence of a reading contract characterising influencers' interactions with their audiences. Varying across influencer classes, in coherence with their fame type, these contracts are the cornerstone upon which influencers' images are based and from which their influence derives. To grow an engaged audience, creators should maintain a fair understanding of how they are perceived as influencers and adopt a publication strategy aligned with this perception. Every new publication then represents a challenge for a creator who puts into play the influence they have developed to test the adequation between the image they believe they have and the image their followers associate them with. We examine two types of feedback here, engagement metrics and user reports, which followers use to respond to unusual content that has breached the reading contracts to various degrees. We observe different types of benign contract breaches (that do not challenge the reading contract) and critical contract breaches (which may permanently change an influencer's image) and discuss their implications in light of anthropological theories.

*The dilemmas of reciprocity*

Tab 4 presents the deviation to the weighted averages of likes and comments per influencer class and type of content for $TIP_2$ and $TIP_3$. A detailed analysis of each class would require a much larger sample of influencers, but this analysis provides general trends.

With regards to $TIP_2$, we observe that content expressing pride for one's achievements (P), self-derision (S) and humour (J) has a positive impact on the number of likes across most classes. This type of content also increases the number of comments for external influencers (especially class 1) but not for organic influencers. By contrast, motivational (M), excited (E) and charity-related content (C) have a negative impact on likes for all classes and on comments for most. Finally, altruistic content (O) decreases engagement for external influencers and increases it for organic ones. Overall, these observations temper those of Burke and Develin (2016), who found that positive feelings, especially 'feeling excited,' trigger more likes and comments than posts without emotion. This may be because their research was conducted on Facebook and focused on a specific feature (poster-annotated feelings); nevertheless, we observe that engagement variations also depend on the influencers' class, and that content displaying excitement tends to trigger lower engagement across all classes than neutral posts.

With regards to $TIP_3$, content outlining recommendations (3R) triggers significantly less engagement across all classes, while content expressing thankfulness to followers (3T) increases it. Furthermore, with the exception of class 3, we observe a clear impact of the communication type on comments: significantly lower number of comments for closed communication (1-2) and unilateral communication when directive (3R), and significantly higher for unilateral communication expressing thankfulness (3T) and open communication (4-5-6).

| | | | Likes — Influencer class | | | | | | | Com. — Influencer class | | | | |
|---|---|---|---|---|---|---|---|---|---|---|---|---|---|---|
| | | 1 | 2 | 3 | 4 | 5 | 6 | | 1 | 2 | 3 | 4 | 5 | 6 |
| Label 2 | H | 0.1 | 0.1 | 0.0 | 0.0 | 0.0 | 0.7 | H | -0.2 | -0.5 | 0.0 | 0.0 | -0.4 | -0.7 |
| | N | -0.2 | -0.1 | 0.1 | -0.1 | -0.2 | 0.0 | N | -0.3 | 0.4 | -0.1 | 0.0 | -0.2 | -0.1 |
| | E | -0.7 | -0.4 | -0.4 | -0.7 | -0.1 | -0.3 | E | -0.2 | -0.6 | -0.4 | -0.8 | 2.1 | -0.1 |
| | P | 0.2 | -0.2 | 0.3 | -0.7 | 0.3 | 0.3 | P | 0.7 | -0.4 | 0.4 | -0.7 | 1.6 | -0.5 |
| | S | 0.9 | 0.4 | -0.1 | 0.4 | 0.3 | 0.3 | S | 0.8 | 0.2 | -0.6 | -0.6 | -0.6 | -0.3 |
| | J | 1.1 | 0.0 | 0.0 | 0.2 | 0.5 | 0.0 | J | 1.6 | 0.6 | 0.2 | -0.6 | -0.4 | -0.7 |
| | L | 0.1 | 0.2 | 0.0 | 0.0 | 0.2 | -0.3 | L | 0.0 | -0.4 | 0.8 | -0.4 | -0.2 | 2.1 |
| | M | -0.5 | -0.2 | -0.6 | 0.1 | -0.1 | -0.1 | M | 0.1 | -0.5 | -0.8 | 1.0 | 0.0 | -0.7 |
| | G | -0.1 | 0.2 | 0.2 | 0.4 | 0.1 | 0.1 | G | 0.6 | -0.3 | 0.1 | 0.6 | -0.2 | -0.2 |
| | O | -0.5 | -0.1 | 0.0 | 0.6 | 0.3 | -0.4 | O | -0.2 | -0.2 | -0.3 | 0.4 | 0.4 | -0.6 |
| | C | -0.8 | -0.3 | -0.6 | -0.5 | 0.0 | -0.8 | C | -0.6 | -0.7 | -0.8 | -0.8 | -0.9 | -1.0 |
| Label 3 | 1 | 0.2 | 0.1 | 0.1 | 0.1 | 0.2 | 0.4 | 1 | -0.1 | -0.2 | -0.3 | 0.0 | -0.5 | -0.3 |
| | 2 | -0.2 | 0.1 | -0.1 | -0.5 | -0.1 | 0.1 | 2 | -0.3 | 0.1 | 0.4 | -0.5 | -0.4 | -0.5 |
| | 3R | -0.6 | -0.6 | -0.4 | -0.4 | -0.6 | -0.2 | 3R | -0.4 | -0.5 | -0.4 | -0.5 | -0.6 | -0.3 |
| | 3W | 0.1 | 0.0 | 0.5 | -0.1 | 0.1 | 0.1 | 3W | 0.0 | 0.0 | 0.0 | -0.2 | -0.3 | -0.5 |
| | 3T | 0.2 | 0.9 | 0.4 | 0.2 | 0.8 | -0.4 | 3T | 0.4 | 0.2 | -0.2 | 1.2 | 0.0 | -0.6 |
| | 4 | 0.2 | 0.0 | -0.5 | 0.4 | 0.1 | -0.2 | 4 | 1.3 | 1.5 | -0.4 | 0.9 | 0.3 | -0.1 |
| | 5 | -0.4 | -0.9 | - | - | 0.2 | -0.2 | 5 | 14.0 | -0.9 | - | - | 23.5 | 18.2 |
| | 6 | 1.0 | -0.7 | -0.2 | -0.1 | 0.6 | 0.6 | 6 | 2.1 | -0.1 | -0.5 | 0.5 | -0.3 | 0.5 |

**Tab 4.** Deviation to the weighted averages of likes and comments per influencer class and type of content for S1 posts[1]

These results allow us to identify three types of benign contract breaches, which impact engagement metrics significantly without challenging the reading contract in the long term.

---

[1] The number of likes of class 7's H posts is not representative as it is strongly impacted by one extreme outlier.

***Communication mistakes:*** Overall, we observe that followers expect influencers to post content associated with the reading contract and do not want to be told what to do. Direct recommendations and charity-related content are unanimously sanctioned by lower engagement metrics across all classes, and motivational and altruistic content have lower engagement metrics for external influencer classes. This suggests that influencers have little moral authority over their followers. Instead, their influence seems to proceed from emotional content (pride, self-derision) when it is in line with the reading contract. Paul Dumouchel understands emotions as 'key moments in the process of affective coordination' whose value dwells on their communication dimension to rally a group of individuals to one's interpretation of a given situation (Dumouchel 1995, 80). According to this perspective, emotional content can 'succeed' or 'fail' to convince others, and it appears that some influencers are more successful at communicating some emotions than others. For instance, class 1 influencers, whose influence revolves around a highly self-centred image, are perceived as less legitimate or authentic when they post altruistic content than class 5 influencers, who are viewed as more 'relatable.' C, O, and M content published by external influencers may then be considered communication mistakes, i.e., mismatches between what the influencer publishes and what followers are willing to listen to.

***Directive frictions:*** Direct recommendations are also associated with lower engagement but are distinct from communication mistakes because they proceed from a challenge inherent to the use of one's influence. The dilemma is the following: while the expected rationale for growing one's influence potential on social media is to be able to use it, influencers tend to breach the reading contract upon which their influence is based when using their influence, i.e., when suggesting a specific behaviour to their audience. In other words, it seems that they cannot use their influence without weakening it. Our observations are consistent with previous research, especially Cheng and Zhang, who identified a 'burning effect' in terms of reputation and engagement when YouTube influencers post sponsored content; the effect's severity increases with the influencers' audience size. They also find that such an effect decreases 'when the sponsored content is more aligned with the influencer's 'usual' content' (Cheng and Zhang 2022, 6), which is also consistent with our hypothesis of a reading contract. Another strategy to overcome this dilemma seems to be the generalisation of contests[2] ($TIP_3 = 5$). Across all classes, we observe that contests log 30x more comments than non-promotional content and 65x more than other types of promotional content, and they also reach the greatest number of likes. Such peaks are not surprising because followers are explicitly asked to engage with the publication to be considered for the lottery. Yet, audiences clearly respond positively and massively to this proposal, offering influencers a valuable alternative to commercials that enables them to benefit from their influence without harming it. People know the importance of brand partnerships for influencers to earn money and improve their content quality. However, traditional promotion content posted by influencers imposes a product on a community's attention, whereas contests internalise sponsors, including them in the relationship between influencers and followers. The friction is lowered because the metamessage is different: followers do not feel objectivated by an influencer trying to sell them a product and incentivised to overrate its quality but appreciate the reciprocity of the interaction, whereby the visibility they give to a brand is rewarded by the possibility of winning a reward.

***Limited positive exceptions:*** As captured by $TIP_3$, followers reward reciprocity. While closed communication posts feign to ignore the audience, open communication posts explicitly recognise followers and value their interactions. Between the two, unilateral communication expresses different types of attention to the audience, from followers perceived as receivers of recommendations (3R) to followers viewed as a supportive community allowing the influencer to grow (3W, 3T). Tab 4 shows that the more open the communication, the higher the engagement. 3T posts, which are quite rare for all classes and trigger some of the greatest levels of engagement – except for class 6, where the relationship with the influencer is highly de-personified and activity-based – particularly demonstrate this reciprocity because posts G, which also express gratefulness but not necessarily directed to their audience, only have a limited impact on engagement. One may then wonder why influencers publish so little open communication content even though it triggers significantly higher engagement metrics. Why, for instance, does class 1 mainly publish showy, love-focused, and neutral content even if they have significantly higher engagement when expressing pride for self-achievements, self-derision, or humour? Why are 89.5% of their posts structured as closed or unilateral communication (3R, 3W) if they recurrently record higher engagement with grateful posts (3T) and open communication? As we understand it, it is that influencers cannot abuse these posts without putting their image at risk due to the tension between reciprocity and desire.

---

[2] The engagement metrics associated with contests in figure 8 for classes 1 and 2 are not relevant because they refer to only one publication each.

The fundamental importance of reciprocity in human interactions has been well-established in psychology and anthropology. It has also been observed online, for instance, by Grinberg et al. (2016), who found that Facebook users increase the consumption of, and interactions with, friends' content before and after posting content; this suggests that, in addition to direct and indirect reciprocity mechanisms ('friends have engaged with my content so I will engage with theirs'), part of this additional activity results from the anticipation of feedback ('I like yours so you will like mine'). This corroborates Kizilcec et al.'s findings on online gifts on Facebook, which show that individuals are 'much more likely to give gifts on their own birthday, whether or not they received a birthday gift.' (Kizilcec et al.'s 2018, 5). This could explain a common strategy of new Instagram users who aim to grow their audience by following many random users and hoping that these will follow them in return. In addition, Scissors et al. observed that Facebook users do not grant all likes the same value, 'desiring feedback most from close friends, romantic partners and family members other than their parents.' (Scissors et al. 2016, 1501). People also certainly value rarer feedback from specific people more, especially feedback from people they admire. Therefore, considering (1) that people tend to value individuals they consider superior and to seek recognition from them, (2) that the amount of collective attention, captured by audience size on Instagram, is a strong signal of such a superiority, (3) that larger audiences mechanically leads to lower individualised reciprocity, and ever to lower generalised reciprocity in some cases (classes 1 and 4), it results a negative correlation between the volume of reciprocal interactions and the value attributed to these interactions, as people tend to seek reciprocity from those who are the least likely to give them.

From an economic viewpoint, an agent who keeps giving to another without any reciprocity would redirect their gifts toward another agent. However, from a psychological perspective, the repeated absence of reciprocity is perceived by the giver as recognition denial, increasing the receiver's prestige and exacerbating the giver's desire to be recognised by them: the influencer's lack of recognition towards me confirms their superiority, justifying my need to keep seeking their validation. Influencers are then incentivised to organise the lack of reciprocity, using the strategy of the coquette that, as we discussed, is very efficient on Instagram. The platform, by design, allows interactions to grow without reciprocity because one does not expect a global influencer with 10+M followers to read and respond to comments. The lack of direct reciprocity is, therefore, not associated with a recognition denial; rather, reciprocity may be expected at the community level. Followers keep interacting with the influencer and growing the influencer's prestige in the hope that this prestige will benefit them at some point when the influencer finally reciprocates and pays tributes to the community by expressing their gratitude. While the strategy of the coquette is to defend a strict autonomy, preventing her from expressing such reciprocity, classes 1 and 5, who loosely follow this strategy, cannot avoid reciprocity because of the competitive nature of Instagram: if followers believe they will never receive recognition, they will likely redirect their attention to a rival influencer. The resulting dilemma for mimetic influencers is that of minimising reciprocity without triggering disengagement or, in other words, maintaining a proper distance from their audience. The more they express this distance from 'common mortals' with self-centred posts and closed communication, the more they fascinate their audience. However, they run the risk of triggering disengagement or hostile feedback if they take this too far. An influencer with too much distance is desired without being loved; they trigger fascination without empathy. Conversely, an influencer too close to their audience develops a sane relationship at the expense of their ability to fascinate.

This explanation helps interpret rare posts, which express unusual emotions or open communication and diverge from the image usually expressed by the influencer, as well as the positive responses of followers to these. These exceptional events act as a regulatory mechanism, a decompression valve allowing the upkeep of the influencer's image. Class 1 are particularly distant from their followers; when they post self-derision or humoristic content, they temporarily reduce this social distance, counterbalancing the inaccessible image they have built by showing that they are capable of not taking themselves too seriously. This interpretation is also supported by the presence of very rare posts expressing sadness, disgust, or authenticity – unusual moments of sincerity which trigger strong engagement (see Tab A.7). Finally, it should be noted that the number of 3W and 3T posts in S1 is inflated by the period of the year in which the data collection was performed (most 3W referring to Christmas and New Year's wishes); these content types are expected to be even less prevalent throughout the full year.

*Social medias reshape the expressions of reciprocity in social interactions*

We now focus on users reporting influencers' accounts to Instagram as a different type of feedback, which is anonymous and invisible to other users. The relative scarcity of these reports prompts us to narrow our analysis to world_stars influencers, who account for 98% of the 33,657 reports considered, and more specifically to class 3 world_stars influencers. Among them, we focus on 'singers' (N = 44), who distinguish themselves by being reported 10x more than 'actors' and 'sportsmen.' Figure 4 presents the distribution of reports against world_stars singers over 60 days in relation to their audience size on a 10-base logarithmic scale. As one may expect, there seems to be a positive correlation between audience size and the number of reports: reports being rare events, it is intuitive that one is more exposed to these when gaining visibility. All singers are American and followed by relatively younger audiences (5.4 years younger on average); furthermore, they tend to be reported by even younger users (6.5 years on average). These trends are amplified for the 25% most reported singers whose names appear on the figure, who are 4.3 years older than their average follower and 8 years older than their typical reporter.

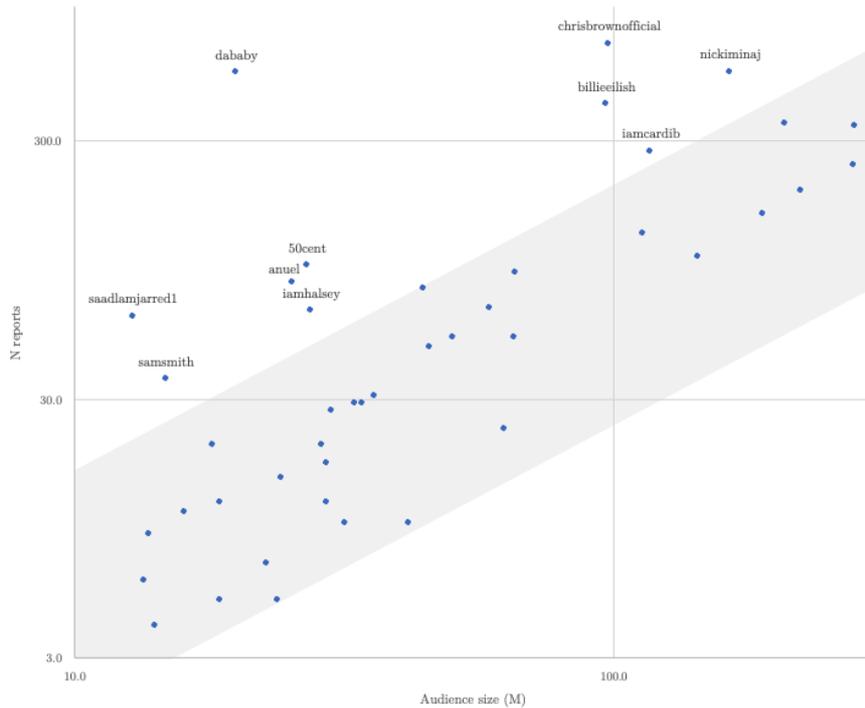

**Figure 4.** Number of reports per audience size for world_stars singers in S0 over 60 days

Two influencers, in particular, catch our attention. Responsible for 65% of all reports against singers and 36% of those against S0 influencers, Travis Scott is an outlier so significant it cannot fit in Figure 4. The 30-year-old singer is followed by an audience of 62% men who are, on average, 5 years younger than him, but he is mainly reported on by women (75%) who are, on average, 7.5 years younger. More specifically, female reporters are close in age to his female followers, but his male reporters tend to be older than his male followers (see Figure A.3). A combination of signals – including when these reports were logged, the tags under which they were submitted (73% of hate or violence tags, whereas 'spam' is the most common tag for user reports), and the fact that the second most reported influencer of our dataset is his partner, Kylie Jenner, with 8% of all reports – allows us to link this abnormally high number of reports to the Astroworld scandal, which cost the life of ten people on November 5$^{th}$, 2021. As the event organiser, Travis Scott was heavily criticised for his communication and perceived lack of empathy following the dramatic event (Da Silva et al. 2021). DaBaby, the second most reported singer of our dataset, with 27.6 reports/M followers, is also reported on by a population quite different from his audience. The 30-year-old rapper is followed by a community composed of 66% men who are, on average, 26 years old, but he is reported by 84% of women with an average age of 22.7 years, and 60% of the report tags relate to hate and bullying. Here again, the combination of report dates and tags, together with the reporters' profiles, allows us to link this reporting to a specific event: a violent altercation between the rapper and his girlfriend DanilLeigh, which was widely reported in the press (BBC 2021).

The findings from previous sections lead us to believe that the degree of content personification would be a meaningful signal to predict the volume of reports against influencers and that the study of user reports associated with influencers' posts (instead of their accounts in general, as we are doing it here) would greatly complement our findings about communication mistakes and contract breaches more generally. Unfortunately, the limited data at our disposal, especially the low number of reports and of singers' posts labelled in S1, do not allow us to conduct further analysis. Yet, the phenomenon observed with Travis Scott and DaBaby suggests meaningful implications for social media studies and promising bridges with philosophical and anthropological theories. Three of these are particularly salient.

First, social media influencers represent an original type of mimetic model with implications for the structure of negative reciprocity, such as rivalry and violence. According to Girard's paradigm, individuals copy their desires from mimetic models they admire and aspire to become similar to. The more they imitate them, the smaller the perceived 'spiritual distance' between them, and the more likely they will develop feelings of rivalry towards them. Through the mimicry of their master, an art student may then rise to their level and, because they do copy not only their style but also their desires, end up competing for the same positions and awards. This dynamic of the mimetic desire, which logically prompts desires to converge and humans to compete, is why Girard believes that we should not imitate human models but God, because the spiritual distance which separates God from us is unsurpassable, leaving no room for rivalry. An interesting aspect of social media influencers is that the spiritual distance that separates them from their followers also tends to be unbridgeable, especially for world celebrities, resulting in an almost null probability of the convergence of desires that could trigger rivalry. Thus, social media influencers represent a kind of model particularly unlikely to lead to a mimetic rivalry between individuals and their mediators that originates violence. Rivalry may still occur but horizontally rather than vertically, akin to courtesans' rivalry around a king – between followers or between influencers, such as class 4 imitating class 1 influencers, or between different communities – exemplified by violent offline clashes between rappers' fans.

Second, the digitalisation of social influencers impacts the type of reciprocity failures and recognition denials which traditionally shape social interactions. The pathologic deviation of the relationship between an individual and their mediator from respectful admiration into hateful rivalry originates from a critical event that Etienne associates with a 'cognitive dissonance' (in review) in the wake of Festinger's theory (1957). The individual experiences what they interpret as a rejection from their mediator – whether accurate because the mediator feels challenged in their mentorship position or inaccurate due to a misunderstanding – resulting in a cognitive dissonance which they resolve by changing the signification associated with their mediator: the inspiring and benevolent model is now perceived as a jealous and hostile rival. As the potential of rivalry between influencers and their followers is all the more limited by the size of the audience and the fact they live in distinct social spaces, the experiences of rejection should be different from those discussed by Girard. Rejection may result from a lack of perceived reciprocity, i.e. from an influencer failing to acknowledge what they owe to their following community, such as Angèle's fans pulling her hair at the end of a concert due to their anger at her lacking time to sign autographs (Rensonnet & Vdh 2021). It may also happen following a 'critical contract breach.' Our analyses reveal significant gaps in engagement when influencers test the bounds of their reading contracts with unusual content, and it is likely that the most severe contract breaches trigger the strongest reactions, from lower engagement to unfriending, hateful comments and reporting. Further research is necessary to investigate the plasticity of influencers and determine the limits that the audiences of each influencer class are ready to accept before assessing an unusual behaviour as too divergent from the influencer's image to maintain a coherent positive representation of them and the types of negative feedback associated with these critical levels. For example, it was estimated that making the cover of *Harper's Bazaar Russia* costed Nicki Minaj up to 1.2 million followers on Twitter and Instagram because of Russia's anti-LGBTQ stances, which were judged as incompatible with the artist's image (Silvia 2018). Logan Paul famously lost hundreds of thousands of YouTube subscribers who were scandalised by his behaviour in his 'suicide forest' video, but the video also attracted 300K new subscribers (Kaplan 2018). Similarly, Travis Scott lost c. 550K followers in the six days following the Astroworld scandal, of which 61% were women who were, on average, 22.5 years old and therefore matched the profile of most of his reporters. However, he also gained 783K followers, balanced between men (49%) and women (51%), with the women being 25.5 years old on average (resp. 219K and 275K for Dababy, see Figure A.2). When influencers reveal traits that are manifestly incompatible with their image, scandals shock their communication systems and reshape influencers' images and audiences. Newly acquired followers may then be either more aligned with the new image or simply curious people attracted by the visibility boost of the scandal who become increasingly interested in the influencers' lives as they start following their activity.

Third, the deviated use of users' reports as a means to sanction an out-of-reach person for an offline behaviour offers a direct reciprocity channel to the mobility of violence. The reporting feature was designed for Instagram users to report policy violations to the platform's administrators and thus trigger moderation actions. However, Etienne and Çelebi (2022) identified that users report accounts for other reasons, including expressing negative feelings about a user and inconveniencing the user, which is consistent with other research (Grossman et al. 2020, Smyrnaios & Papaevangelou 2020). Considering that neither Travis Scott nor DaBaby's accounts broke Instagram's policies during the considered period, the exceptionally numerous reports logged against the two rappers likely belong to this category and can be interpreted as an attempt to sanction them for their offline behaviour. Following McLuhan's conceptions of media as instruments extending people's selves and senses, this sanctioning suggests that followers view Instagram accounts as an extension of the influencers, and thus as a way to react to behaviours unrelated to their online activity. As influencers are out of reach in the real world, people angry at them seek to reach them via their Instagram accounts. When experiencing violence but being unable to reciprocate it directly, which occurs especially when its author is out of reach, people tend to take revenge by transferring their violence to another person who is arbitrarily selected but within reach. This shift from direct to indirect reciprocity is known as the 'mobility of violence' or 'displaced aggression' (Dollard et al. 1939), whose robustness was confirmed by a significant body of experimental literature (Marcus-Newhall et al. 2000). Anspach mobilises this concept to support Girard's scapegoat mechanism (Anspach 2011), according to which human societies, when fragilized by internal conflicts, re-cement themselves by rejecting one individual against whom individuals' hate polarises and who then becomes the target of their collective violence (Girard 1972). Building upon these theories, our observations suggest that digital media do not simply reproduce real-world interactions but truly extend them, as illustrated by Instagram, which offers a direct reciprocity channel to the mobility of violence in the form of reporting, which could contribute to reducing the need for indirect reciprocity in the real world.

Social media produces interactional spaces where mimetic rivalry is less likely to develop between mimetic models and their followers because of the distance established by the digital platform's design. However, it also results in relationships that tend to be more unstable than those discussed by Girard because this distance encourages individuals, when constituting themselves as influencers, to shape their image in the wake of the feedback they receive from the community. This exposes them to the permanent risk that their actions will diverge too much from their expected behaviours, provoking a dissonance likely to turn their mimetic followers' positive feelings into negative ones. The greater the fascination, the most likely the attention turns into hate because the actual model can only disappoint compared to their idealised representation. Influencers build communities by distinguishing themselves within attention spaces as mimetic models; by the same token, they also self-designate as the potential target of their hate. Describing the process through which leaders originally emerged in archaic societies, Girard explains that they were always 'victims awaiting sacrifice' (Girard 1972, 161). Such is a teaching that influencers should not forget: they always run the risk that, following a scandal impacting their image, the community of admirers formed under their figurehead ends up recovering their lost unity by turning their positive idol into a collective victim, a common target against whom their disappointment can converge in a hateful manner. Such is the process through which social groups have always dealt with conflicts, according to Girard, totalising the community against a common enemy whose sacrifice allows a collective catharsis.

## Conclusion

By approaching social network influencers as communication systems, we aimed to expose a sophisticated approach to interactions between influencers and their followers. We proposed two original classifications which allowed us to distinguish between six classes of influencers, characterised by different audiences and organised around reading contracts aligned with the type and origin of the creator's fame. The growing portion of personal content posted by influencers as their audiences grew then prompted us to characterise 'real' general influence, which is independent of the publication's topics, highly person-based and derives from the Girardian mimetic desire. Three influence growth strategies were then observed: the keepers, who satisfy themselves with a specialised influence whose potential is limited, the shifters, who intend to move from an activity-based to a person-based influence at the risk of breaching their reading contract, and a third group who plays the Girardian coquette strategy, which Instagram's design seems to reward greatly in terms of engagement and brand sponsorships. Finally, we identified four types of reading contract breaches associated with various follower feedback in terms of engagement metrics and user reports and discussed their implications for social sciences' theories about social mimicry, reciprocity and violence.

As a foundational study, this paper does not aim to present rigorous demonstrations of statistically significant correlations. The great number of variables at play in such complex social interactions, the imperfect metrics at our disposal to capture them, and the limited data analysed for this research should refrain anyone from exaggerating our conclusions. Nevertheless, we believe that this paper presents a coherent conceptual framework and a valuable methodology to study influencer-centred online interactions, illustrated by quantitative results. It also lays a valuable milestone towards more cross-disciplinary research, bringing closer quantitative social media studies and philosophical anthropology. The digitalisation of social interactions impacts the expressions of social comparison, mimetic behaviours and reciprocity as much as it creates room for improvement. By changing the design of social platforms, one can promote specific reciprocity types versus others. By better understanding followers' feedback, influencers can improve their communication to reduce the occasions of dissonance and the experiences of reciprocity failure.

## Ethical statement

The experimental protocol was approved by Meta's internal review board, which ensured that all methods were carried out in accordance with relevant guidelines and regulations.

## Data availability

All data analysed in the scope of this study are included in this published article within the associated dataset.xlsx file. To comply with existing regulations on users' privacy, the IDs of Instagram users present in S0 were anonymized, and the images associated with their publications in S1 were removed.

Annexes

## I. Annotation methodology

*The classification of influencers per fame*

The classification of influencers per fame (CIF) contains four verticals. Creators' activities were collected via internet research, and when an influencer cumulated different activities along their career (e.g., a sportsman becomes a TV presenter), the activity selected is that which truly made them famous to the general public. Among class 3 influencers represented in S0, sportsmen include football players, American football players, basketball players, cricketers, Formula 1 runners, boxers, skateboarders, pole vaulters, bull riders, dancers and ice dancers. The type of fame is associated with the individual's activity: it is activity-based when the individual's performance can be dissociated from their personhood (e.g., one may enjoy Booba's music without appreciating the singer as a person and consume their music without being interested in their private life) or person-based when such a dissociation tends to become impossible.

Unlike external influencers, organic influencers grew their fame as content creators on social media. Their activity is, therefore, closely related to the kind of content they produce on Instagram, YouTube or Tiktok, although they may have another professional activity irrelevant to their influencer status and often difficult to identify – especially for class 6, whose creators have an almost anonymous and mediatic presence often exclusively limited to Instagram. As a result, while the degree of content personification appearing in the CIF results from the observations presented in section 2, relevant to external influencers, it is more linked to the degree of content personification for organic influencers per the design of the classification. Class 4 contains pure social media players (e.g., bloggers born on Instagram or Tiktok) who mainly showcase their bodies. Class 5 contains pure social media players (e.g., family blogs, travel blogs, pages of YouTubers or Tiktokers who publish different types of content on their primary social media), coaches (sports, food, etc.) and passion broadcasters (wine, hunting, cars, etc.). Class 6 contains artists (tattoo, interior design, clothes, jewellery, flowers, etc.) and passion broadcasters (fishing, memes, motorcycles, cars, dogs, etc.)

*The typology of influencers' posts*

The typology of influencers' posts (TIP) has three labelling levels.

The first level describes the publication's pictorial content. It labels who is present in the post: 'Me' (the influencer alone, or eventually with their baby or young children), 'Me & others' (the influencer surrounded by other people), 'others' (only other people) or 'nobody' (the content only contains an object, a landscape, a screenshot of another post, etc.). It also distinguishes between personal content (1, 2, 3), associated with the influencer's private life, and professional content (4, 5, 6), associated with the influencer's activity. Hence, the 'other' people tend to be friends and family members for posts labelled 2 or 3, but colleagues, clients and professional contacts for posts labelled 5 or 6. No distinction was made for label 7 between personal and professional life, and posts only presenting influencers' children or pets were rated 2 as we consider them extensions of their parents/owners.

The second level captures the holistic and general emotions that emanate from the post by combining both images and captions. It intends to account for different ways of presenting a given message, thus informing about how an influencer wants to be perceived when communicating. Our 14 categories include H for showy (influencers strike a serious pose without smiling to look inaccessible), E for excitement (usually related to an announcement such as a pregnancy or the release of an album, a movie, an article, etc.), P for pride (related to achievements), S for self-derision (posts where influencers are not presented at their best, mocking themselves or expressing weaknesses), J for humour, L for love (expressing the influencer's attachment of goodwill vis-à-vis their family, friends or colleagues), M for motivational (spiritual or motivational claims to provide advice, support and encourage followers), G for gratefulness (influencers expresses their gratitude to specific people or, in general, for what they have), O for altruism (celebrating another person for their work or friendship), C for charity (explicit supporting a charity cause, often encouraging donations), X for sadness (expressing compassion, sadness or bereavement), D for disgusted (positioned statements when influencers express anger or disgust regarding a situation), A for authenticity (influencers being open and genuine about their failure and life difficulties) and N for neutral. As several of the categories can apply to a unique post, $TIP_2$ can combine several letters (e.g. 'AGL').

The third level captures the type of communication influencers set up with their posts. Every post is viewed as a message from an influencer to their community – otherwise, one would wish their friends a happy birthday in private rather than through public posts – but the way this message is structured suggests different kinds of responses. The communication can be entirely closed on the influencers' self: type 1 is for posts where influencers only discuss their lives as if they were talking to themselves. It can also be relatively close: type 2 refers to posts where influencers are talking to a specific person as they would do in a private message ('I am so blessed to have you X, as a friend') or discuss something largely unrelated to themselves (e.g. a picture of a landscape with no caption). The communication can also be unilateral, which is the case when influencers directly address their community without expecting a response: type 3R for recommendations ('I suggest you check out this artist' page,' 'I recommend this product,' 'go check out my last show'), 3W for wishes ('I wish you a great day,' 'a Happy New Year', 'a Merry Christmas') and 3T for thanks ('thank you so much for your support,' 'Your love and comments encourage me to continue'). Lastly, the communication can be open, calling for followers' engagement: type 4 for direct questions ('What do you think about my new look?', 'tag a friend if you agree with this,' 'reach out by direct messages if you want to buy this piece of art'), type 5 for contests ('I am partnering with brand X to offer you this trip to Italy. Like and comment the post with your name to participate and the winner will be drawn at random') or type 6 (replies to followers' comments, live Q&A, meeting with followers at events such as book signings).

The labelling task was performed by the researchers, which should not be considered a limitation of the methodology for at least three reasons. First, because while outsourcing the labelling task to external annotators is relevant to label voluminous databases with simplistic labels (e.g. is there a plane on this picture?), it becomes irrelevant to analyse social interactions with more sophisticated labels. Second, because the annotation task was conducted with the utmost seriousness, including multiple rounds of reviews. Third, because the typology was designed in such a way that the task can be replicated. Labels 1 and 3 are based on the structure of the content – the components of the pictorial content for the former, the semantic structure of the latter – which drastically limits the weight of annotators' subjective appreciation in the annotation task. Label 2 is arguably more subjective, which is why the researchers mainly relied on keywords – e.g. almost all G-rated items' captions include words such as 'thanks,' 'cheers,' 'grateful,' 'honoured,' 'blessed.'

## II. Presentation of the influencers' accounts dataset (S0)

S0 is composed of a random selection of N = 619 public Instagram accounts balanced across four regional groups: France, the US, the UK and 'world_stars.' This latter group gathers all accounts from S0 whose audience exceeds 10M followers, whether they are owned by French, American, or British influencers. Tab A.1 presents the resulting composition of S0 per influencer class and region. All data was extracted from Instagram internal servers on the 31st of October 2021 and is available under an anonymised format in the file dataset.xlsx.

| Influencer class | 1 | 2 | 3 | 4 | 5 | 6 | *N* |
|---|---|---|---|---|---|---|---|
| World_stars | 9 | 9 | 118 | 7 | 8 | 1 | *152* |
| US | 10 | 23 | 31 | 26 | 29 | 37 | *156* |
| FR | 16 | 24 | 27 | 32 | 35 | 20 | *154* |
| UK | 49 | 23 | 25 | 15 | 21 | 24 | *157* |
| *Sum* | *84* | *79* | *201* | *80* | *93* | *82* | *619* |

**Tab A.1.** Composition of S0 per influencers' class and region

To ensure our analysis was not narrowed to 'mega-influencers,' we also aimed to select a critical number of accounts from different audience ranges and filtered both private accounts (for privacy reasons and because influencers' accounts are typically public) and non-personified accounts held by organisations (e.g. National Geographic) or brands (e.g. Dior). Tab A.2 presents the resulting composition of S0 per audience size and region and estimates S0's coverage *vis-à-vis* the population relevant to our study – e.g., the 34 British accounts with a 10-50M follower base in S0 account for 77.5% of all UK public and personified influencers accounts followed by 10K to 50K followers on October 31st, 2021.

As shown, S0 covers a significant part of all accounts followed by more than 10M people in the three countries, and 36 out of 40 of the influencers followed by more than 50M people while still including micro- and nano-influencers. Therefore, if one argues that the world_stars category does not include enough instances to allow for a rigorous analysis (with the exception of class 3), we shall answer that it contains almost all the mega-influencers, representing dozens of millions of followers.

| Country / Audience size | 10-50K | 50-100K | 100K-1M | 1-10M | 10-50M | >50M |
|---|---|---|---|---|---|---|
| US | 38 | 33 | 46 | 39 | 70 | 34 |
| UK | 32 | 23 | 47 | 56 | 34 | 1 |
| FR | 32 | 35 | 50 | 36 | 12 | 1 |
| *Coverage* | | | | | | |
| US | 0.0% | 0.1% | 0.1% | 0.8% | 22.9% | 34/36 |
| UK | 0.1% | 0.3% | 0.6% | 6.6% | 77.5% | 1/2 |
| FR | 0.1% | 0.7% | 1.3% | 11.1% | 100.0% | 1/2 |

**Tab A.2.** Composition of S0 per audience size and region

| Influencer class | 1 | 2 | | 3 | | | | | 4 | 5 | 6 |
|---|---|---|---|---|---|---|---|---|---|---|---|
| Subclass | Reality TV | Model | TV presenter | Actor/Singer | Singer | Actor | Sportman | Other | All | All | All |
| World_stars | 130.4 | 40.8 | - | 76.8 | 65.8 | 41.9 | 25.6 | 22.4 | 22.5 | 22.3 | 10.4 |
| US | 2.5 | 2.8 | 1.5 | - | 2.6 | 0.9 | 0.5 | - | 0.4 | 1.2 | 0.2 |
| FR | 2.9 | - | 0.4 | - | 1.0 | 0.9 | 4.5 | 0.3 | 0.2 | 1.3 | 0.1 |
| UK | 1.4 | 0.1 | 1.4 | - | 0.4 | 0.4 | 1.7 | 0.3 | 0.1 | 1.1 | 0.1 |
| Average (US,FR,UK) | 1.8 | 1.2 | | 1.2 | | | | | 0.2 | 1.2 | 0.2 |

**Tab A.3** Average audiences of S0 influencers per class and region in M followers[3]

---

[3] The audiences of subclass Sportman/FR (4.5) and class 6/world_stars (10.4) are not representative because they refer to two and one instances, respectively.

## III. Ages and genders

To confirm the robustness of our results on gender analysis, we observe that there was no significant change in the distributions of genders at the influencer and country levels in a two-month period (from 31/10/2021 to 01/04/2022). At the influencer level: Avg(delta) = 0.28%, Med(delta) = 0.17% and Max(delta) = 2.68%. At the region level: Delta(World) = 0.11%, Delta(US) = 0.86%, Delta(UK) = 0.04%, Delta(FR) = 0.03%. In addition, while all users do not provide their gender, resulting in gaps in gender data, two signals allow us to assume that the gender ratios remain stable between known and unknown genders. First, the proportion of known genders largely outnumbers that of unknown genders (70-78% across countries). Second, these ratios remain stable when distributed across all ages for each of the four regional categories. Finally, while Instagram allows users to select between 'male,' 'female' and 'custom' when choosing a gender, we neglect this latter category, which only represents 0.3% of S0's audiences.

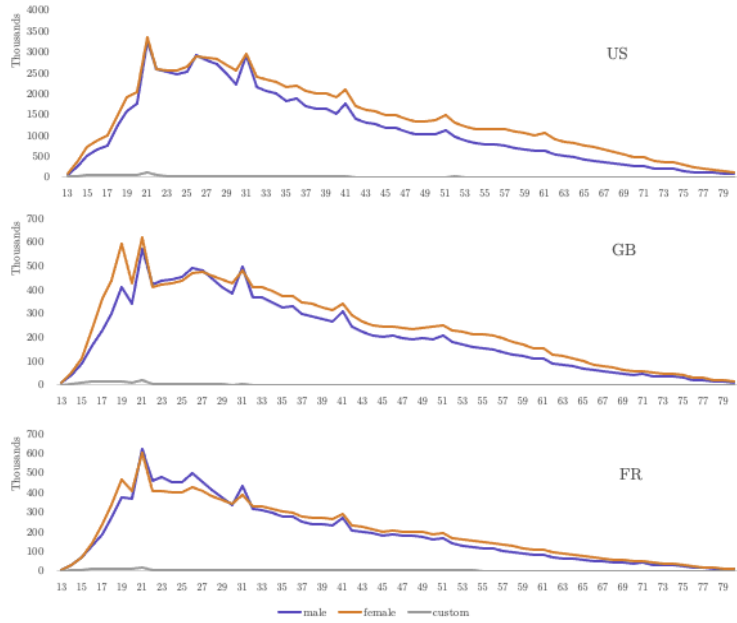

**Figure A.1.** Distributions of Instagram users' genders per age in the US, UK and France

| Class 1 | 13-18 | 19-24 | 25-30 | 31-40 | 41-55 | 55+ | N |
|---|---|---|---|---|---|---|---|
| 13-18 | | | | | | | 0 |
| 19-24 | 147 | 144 | 125 | 81 | 41 | 19 | 7 |
| 25-30 | 128 | 134 | 134 | 94 | 45 | 18 | 31 |
| 31-40 | 116 | 116 | 134 | 114 | 58 | 22 | 36 |
| 41-55 | 132 | 127 | 126 | 98 | 53 | 24 | 9 |
| 55+ | 111 | 119 | 128 | 111 | 64 | 27 | 1 |

| Class 2 | 13-18 | 19-24 | 25-30 | 31-40 | 41-55 | 55+ | N |
|---|---|---|---|---|---|---|---|
| 13-18 | | | | | | | 0 |
| 19-24 | 100 | 125 | 141 | 110 | 57 | 18 | 2 |
| 25-30 | 111 | 138 | 143 | 94 | 45 | 16 | 7 |
| 31-40 | 83 | 106 | 148 | 127 | 66 | 25 | 20 |
| 41-55 | 44 | 59 | 111 | 160 | 140 | 72 | 29 |
| 55+ | 36 | 53 | 100 | 155 | 157 | 98 | 21 |

| Class 3 | 13-18 | 19-24 | 25-30 | 31-40 | 41-55 | 55+ | N |
|---|---|---|---|---|---|---|---|
| 13-18 | 242 | 163 | 83 | 48 | 26 | 7 | 1 |
| 19-24 | 166 | 158 | 118 | 68 | 31 | 11 | 19 |
| 25-30 | 145 | 148 | 129 | 78 | 37 | 15 | 59 |
| 31-40 | 111 | 127 | 141 | 103 | 51 | 20 | 71 |
| 41-55 | 96 | 114 | 138 | 121 | 65 | 21 | 41 |
| 55+ | 77 | 102 | 129 | 125 | 87 | 47 | 8 |

| Class 4 | 13-18 | 19-24 | 25-30 | 31-40 | 41-55 | 55+ | N |
|---|---|---|---|---|---|---|---|
| 13-18 | 268 | 164 | 63 | 43 | 28 | 12 | 11 |
| 19-24 | 219 | 165 | 85 | 52 | 31 | 12 | 19 |
| 25-30 | 75 | 115 | 148 | 122 | 68 | 21 | 24 |
| 31-40 | 28 | 108 | 199 | 139 | 48 | 11 | 16 |
| 41-55 | 4 | 11 | 60 | 175 | 211 | 189 | 2 |
| 55+ | | | | | | | 0 |

| Class 5 | 13-18 | 19-24 | 25-30 | 31-40 | 41-55 | 55+ | N |
|---|---|---|---|---|---|---|---|
| 13-18 | 227 | 186 | 77 | 38 | 21 | 7 | 3 |
| 19-24 | 230 | 167 | 79 | 49 | 30 | 11 | 9 |
| 25-30 | 160 | 166 | 115 | 62 | 32 | 12 | 28 |
| 31-40 | 128 | 147 | 130 | 87 | 41 | 13 | 34 |
| 41-55 | 85 | 113 | 154 | 123 | 57 | 19 | 7 |
| 55+ | 126 | 161 | 124 | 68 | 49 | 18 | 2 |

| Class 6 | 13-18 | 19-24 | 25-30 | 31-40 | 41-55 | 55+ | N |
|---|---|---|---|---|---|---|---|
| 13-18 | | | | | | | 0 |
| 19-24 | 107 | 168 | 162 | 64 | 24 | 7 | 4 |
| 25-30 | 35 | 79 | 159 | 157 | 94 | 35 | 9 |
| 31-40 | 88 | 94 | 138 | 140 | 78 | 21 | 31 |
| 41-55 | 25 | 39 | 111 | 180 | 155 | 83 | 9 |
| 55+ | 2 | 7 | 34 | 81 | 168 | 512 | 2 |

**Tab A.4.** Covariance matrixes of affinity scores for influencers and audience' ages

| Influencer class | 1 | | 2 | | 3 | | 4 | | 5 | | 6 | |
|---|---|---|---|---|---|---|---|---|---|---|---|---|
| | male | female | male | female | male | female | male | female | male | female | male | female |
| World_stars | 2 | 7 | | | 9 | 72 | 46 | 1 | 6 | 5 | 3 | |
| male | 40% | 42% | | | 40% | 62% | 41% | 89% | 61% | 53% | 36% | |
| female | 60% | 58% | | | 60% | 37% | 59% | 11% | 38% | 47% | 63% | |
| US | 1 | 9 | 10 | 13 | 23 | 8 | 2 | 24 | 11 | 18 | 12 | 22 |
| male | 20% | 21% | 40% | 43% | 68% | 26% | 19% | 52% | 75% | 20% | 71% | 21% |
| female | 80% | 79% | 60% | 57% | 31% | 73% | 81% | 48% | 24% | 80% | 28% | 78% |
| FR | 3 | 13 | 15 | 9 | 17 | 10 | 7 | 25 | 22 | 13 | 3 | 15 |
| male | 42% | 23% | 42% | 46% | 54% | 27% | 17% | 33% | 64% | 26% | 78% | 11% |
| female | 57% | 77% | 58% | 54% | 46% | 72% | 82% | 66% | 35% | 74% | 20% | 88% |
| GB | 12 | 37 | 9 | 14 | 14 | 11 | 2 | 13 | 14 | 7 | 6 | 17 |
| male | 23% | 20% | 28% | 28% | 49% | 23% | 27% | 45% | 51% | 9% | 30% | 10% |
| female | 77% | 80% | 71% | 72% | 50% | 76% | 73% | 55% | 48% | 91% | 70% | 90% |
| (US,FR,GB) | 16 | 59 | 34 | 36 | 54 | 29 | 11 | 62 | 47 | 38 | 21 | 54 |
| male | 28% | 21% | 37% | 39% | 57% | 26% | 21% | 43% | 63% | 18% | 60% | 14% |
| female | 71% | 79% | 63% | 61% | 42% | 74% | 79% | 56% | 36% | 81% | 39% | 86% |

**Tab A.5.** Audiences' genders split per influencer class and gender

**Tab A.6.** Audiences' genders split per influencer subclass and gender for classes 3 and 6

| Class | | | 3 | | | | | | | 5 | | | | | | | |
|---|---|---|---|---|---|---|---|---|---|---|---|---|---|---|---|---|---|
| Subclass | Singer | | Actor/Singer | | Actor | | Sportman | | Other | | SM Content creator | | Coach | | Artist | | Passion broadcaster | |
| | male | female | male | female | male | female | male | female | male | female | male | female | male | female | male | female | male | female |
| World_stars | 23 | 21 | 1 | 6 | 16 | 18 | 28 | 1 | 4 | | 5 | 2 | | | 1 | | | |
| male | 49% | 40% | 47% | 36% | 52% | 40% | 82% | 90% | 47% | | 53% | 37% | | | 34% | | | |
| female | 51% | 59% | 52% | 63% | 48% | 60% | 18% | 10% | 52% | | 47% | 62% | | | 65% | | | |
| US | 13 | 3 | | | 4 | 4 | 6 | 1 | | | 7 | 10 | | 7 | 1 | 1 | 3 | |
| male | 65% | 29% | | | 61% | 27% | 79% | 12% | | | 74% | 24% | | 15% | 31% | 9% | 90% | |
| female | 34% | 70% | | | 38% | 72% | 21% | 88% | | | 25% | 75% | | 85% | 69% | 91% | 9% | |
| FR | 5 | 2 | | | 5 | 6 | 2 | 0 | 5 | 2 | 19 | 5 | 1 | 2 | 1 | 3 | 1 | 3 |
| male | 60% | 29% | | | 45% | 24% | 75% | | 48% | 37% | 64% | 32% | 34% | 12% | 68% | 13% | 89% | 37% |
| female | 39% | 71% | | | 55% | 76% | 24% | | 51% | 62% | 35% | 68% | 65% | 87% | 29% | 86% | 11% | 63% |
| male | 63% | 30% | | | 10% | 23% | 55% | | 53% | 18% | 53% | 11% | 26% | 17% | 49% | 6% | 70% | 9% |
| female | 36% | 70% | | | 90% | 76% | 45% | | 47% | 82% | 46% | 89% | 74% | 83% | 51% | 94% | 29% | 90% |
| (US,FR,GB) | 20 | 7 | | | 11 | 17 | 16 | 1 | 7 | 4 | 35 | 16 | 3 | 10 | 3 | 8 | 6 | 4 |
| male | 63% | 29% | | | 38% | 25% | 70% | 12% | 51% | 27% | 64% | 22% | 30% | 15% | 49% | 9% | 83% | 23% |
| female | 36% | 70% | | | 61% | 75% | 30% | 88% | 49% | 72% | 35% | 77% | 70% | 85% | 50% | 90% | 16% | 76% |

## IV. Presentation of the influencers' posts dataset (S1)

S1 is composed of 2,404 content pieces selected among all posts from 204 influencers of S0 between the 31$^{st}$ of October 2021 and the 4$^{th}$ of January 2022. A content piece is composed of pictorial content (one or several images and videos) and semantic content (textual caption). The sampling was randomly stratified, and we tried to have in S1 as many different influencers from S0 as possible, with a target number of 10 posts per influencer, while keeping a balanced aggregate between the six influencers classes of CIF. We also selected influencers who had the most local audience (except for world_stars). The annotation is based on the TIP, which aggregates three levels of labels. Leaving aside sponsored posts, we are left with 2,142 items balanced between regions ($N_{world\_stars}$ = 551, $N_{US}$ = 560, $N_{UK}$ = 518, $N_{FR}$ = 513) and classes ($N_1$ = 378, $N_2$ = 307, $N_3$ = 407, $N_4$ = 345, $N_5$ = 389, $N_6$ = 316).

Tab A.7 presents the distribution of non-promotional content per influencer class and content type for the three labels of the TIP (N), together with their weighted average numbers of likes (likes) and comments (com.), adjusted by influencers' average audience size over the period (/K followers).

|  |  | Influencer class |  |  |  |  |  |  |  |  |  |  |  |  |  |  |  |  |  |  |
|---|---|---|---|---|---|---|---|---|---|---|---|---|---|---|---|---|---|---|---|---|---|
|  | N | 1 | 2 | 3 | 4 | 5 | 6 | Likes | 1 | 2 | 3 | 4 | 5 | 6 | Com. | 1 | 2 | 3 | 4 | 5 | 6 |
| Content Label 1 | 1 | 106 | 57 | 69 | 212 | 85 | 14 | 1 | 32 | 26 | 36 | 67 | 38 | 15 | 1 | 0.3 | 0.5 | 0.5 | 0.9 | 0.5 | 0.2 |
|  | 2 | 152 | 80 | 76 | 89 | 74 | 30 | 2 | 40 | 36 | 38 | 57 | 34 | 28 | 2 | 0.2 | 0.5 | 0.3 | 0.6 | 0.7 | 1.1 |
|  | 3 | 24 | 15 | 17 | 3 | 4 | 3 | 3 | 19 | 19 | 47 | 46 | 37 | 13 | 3 | 0.1 | 0.4 | 2.1 | 1.2 | 0.3 | 0.1 |
|  | 4 | 37 | 54 | 66 | 15 | 51 | 11 | 4 | 15 | 21 | 52 | 62 | 38 | 18 | 4 | 0.1 | 0.3 | 1.1 | 0.9 | 2.3 | 0.3 |
|  | 5 | 31 | 44 | 136 | 2 | 47 | 16 | 5 | 17 | 19 | 45 | 59 | 46 | 21 | 5 | 0.3 | 0.7 | 0.6 | 0.9 | 1.2 | 0.5 |
|  | 6 | 4 | 6 | 2 | 0 | 23 | 3 | 6 | 8 | 50 | 54 |  | 9 | 3 | 6 | 0.1 | 0.5 | 0.6 |  | 0.3 | 0.1 |
|  | 7 | 24 | 51 | 41 | 24 | 105 | 239 | 7 | 6 | 20 | 16 | 15 | 21 | 15 | 7 | 0.1 | 0.9 | 0.2 | 0.3 | 0.7 | 1.0 |
|  | N | 1 | 2 | 3 | 4 | 5 | 6 | Likes | 1 | 2 | 3 | 4 | 5 | 6 | Com. | 1 | 2 | 3 | 4 | 5 | 6 |
| Content Label 2 | H | 104 | 32 | 46 | 170 | 51 | 4 | H | 35 | 31 | 40 | 63 | 35 | 29 | H | 0.2 | 0.3 | 0.6 | 0.8 | 0.6 | 0.3 |
|  | N | 130 | 123 | 148 | 129 | 189 | 228 | N | 26 | 25 | 44 | 55 | 28 | 17 | N | 0.2 | 0.7 | 0.5 | 0.8 | 0.7 | 0.8 |
|  | E | 35 | 24 | 51 | 4 | 31 | 8 | E | 11 | 17 | 23 | 20 | 30 | 13 | E | 0.2 | 0.2 | 0.3 | 0.2 | 2.8 | 0.8 |
|  | P | 16 | 10 | 31 | 4 | 18 | 9 | P | 37 | 21 | 52 | 21 | 45 | 22 | P | 0.4 | 0.3 | 0.7 | 0.3 | 2.4 | 0.4 |
|  | S | 6 | 13 | 10 | 4 | 14 | 3 | S | 59 | 39 | 35 | 83 | 45 | 22 | S | 0.4 | 0.6 | 0.2 | 0.3 | 0.3 | 0.6 |
|  | J | 28 | 37 | 35 | 10 | 29 | 2 | J | 64 | 28 | 39 | 72 | 50 | 17 | J | 0.6 | 0.8 | 0.7 | 0.4 | 0.6 | 0.3 |
|  | L | 79 | 39 | 57 | 25 | 43 | 29 | L | 36 | 33 | 38 | 59 | 41 | 11 | L | 0.2 | 0.3 | 0.9 | 0.5 | 0.7 | 2.8 |
|  | M | 8 | 7 | 15 | 8 | 30 | 6 | M | 16 | 20 | 17 | 65 | 29 | 15 | M | 0.3 | 0.2 | 0.1 | 1.6 | 0.9 | 0.3 |
|  | G | 40 | 44 | 72 | 21 | 33 | 37 | G | 29 | 33 | 46 | 89 | 39 | 19 | G | 0.4 | 0.4 | 0.6 | 1.3 | 0.7 | 0.7 |
|  | O | 13 | 31 | 30 | 2 | 2 | 8 | O | 14 | 24 | 41 | 98 | 42 | 11 | O | 0.2 | 0.4 | 0.4 | 1.1 | 1.2 | 0.4 |
|  | C | 5 | 9 | 11 | 1 | 3 | 4 | C | 7 | 20 | 14 | 34 | 34 | 3 | C | 0.1 | 0.2 | 0.1 | 0.1 | 0.1 | 0.0 |
|  | X | 2 | 7 | 3 | 1 | 1 | 0 | X | 69 | 21 | 11 | 266 | 63 |  | X | 0.4 | 0.3 | 0.3 | 1.2 | 3.1 |  |
|  | D | 0 | 1 | 2 | 1 | 0 | 0 | D |  | 10 | 150 | 77 |  |  | D |  | 0.5 | 16.7 | 4.6 |  |  |
|  | A | 1 | 0 | 1 | 4 | 4 | 0 | A | 145 |  | 55 | 79 | 90 |  | A | 4.0 |  | 0.5 | 4.2 | 4.9 |  |
|  | N | 1 | 2 | 3 | 4 | 5 | 6 | Likes | 1 | 2 | 3 | 4 | 5 | 6 | Com. | 1 | 2 | 3 | 4 | 5 | 6 |
| Content Label 3 | 1 | 148 | 78 | 117 | 210 | 111 | 19 | 1 | 34 | 29 | 44 | 66 | 38 | 24 | 1 | 0.2 | 0.4 | 0.5 | 0.8 | 0.5 | 0.6 |
|  | 2 | 89 | 123 | 160 | 35 | 88 | 117 | 2 | 23 | 28 | 38 | 30 | 28 | 19 | 2 | 0.1 | 0.6 | 0.9 | 0.4 | 0.5 | 0.5 |
|  | 3R | 36 | 35 | 48 | 12 | 58 | 77 | 3R | 13 | 11 | 25 | 35 | 11 | 13 | 3R | 0.1 | 0.2 | 0.4 | 0.4 | 0.4 | 0.7 |
|  | 3W | 51 | 32 | 26 | 37 | 28 | 25 | 3W | 33 | 27 | 63 | 55 | 33 | 19 | 3W | 0.2 | 0.5 | 0.6 | 0.6 | 0.6 | 0.5 |
|  | 3T | 9 | 7 | 28 | 11 | 9 | 14 | 3T | 34 | 50 | 58 | 75 | 56 | 11 | 3T | 0.3 | 0.6 | 0.5 | 1.7 | 0.9 | 0.4 |
|  | 4 | 26 | 15 | 14 | 22 | 61 | 36 | 4 | 34 | 25 | 19 | 87 | 34 | 14 | 4 | 0.5 | 1.3 | 0.4 | 1.5 | 1.2 | 0.8 |
|  | 5 | 1 | 1 | 0 | 0 | 5 | 6 | 5 | 18 | 1 |  |  | 36 | 13 | 5 | 3.4 | 0.0 | 0.0 | 0.0 | 21.9 | 18.0 |
|  | 6 | 2 | 4 | 5 | 2 | 8 | 5 | 6 | 59 | 7 | 33 | 56 | 50 | 28 | 6 | 0.7 | 0.5 | 0.3 | 1.2 | 0.6 | 1.4 |

**Tab A.7.** Summary of content type and engagement per TIP label and influencer class

**IV. User reports against Travis Scott and DaBaby**

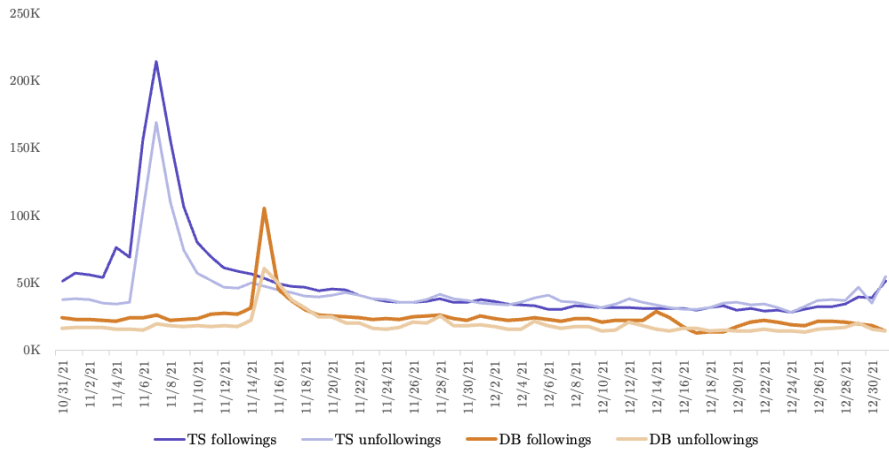

**Figure A.2.** Travis Scott and DaBaby's daily following evolution

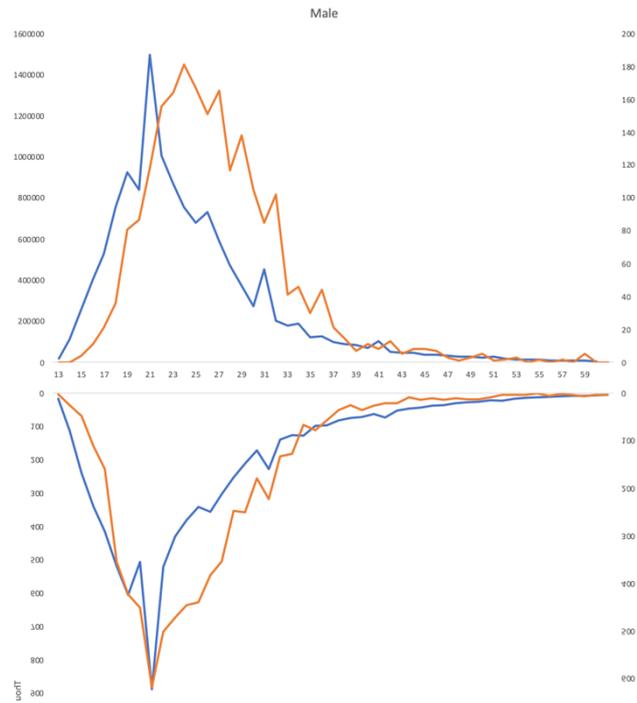

**Figure A.3.** Travis Scott's distributions of reporters and followers per age and gender